\documentclass[a4paper, 11pt]{article}
\usepackage{graphics}
\usepackage{epsfig}
\usepackage{amsmath}

\def\sumint{\hbox{$\sum$}\!\!\!\!\!\!\int}

\newcommand{\measure}[1]{\mathrm{d}#1\,}
\renewcommand{\eqref}[1]{Eq$.$ (\ref{#1})}  		
\newcommand{\secref}[1]{Sec$.$ \ref{#1}}  		
\newcommand{\figref}[1]{Fig$.$ (\ref{#1})}  		
\newcommand{\citerefs}[1]{Refs$.$ \cite{#1}}  	
\newcommand{\spaceint}[2]				
{
 \frac{\mathrm{d}^{#1}{#2}}{(2 \pi)^{#1}}
}	
	
\newcommand{\ie}{{\it i.e.~}}

\begin{document}

\title{\bf{Heating the $O(N)$ nonlinear sigma 
model\footnote{Seminar talk given
at the 43st Cracow School of Theoretical Physics, 30 May - 8 June
2003, Zakopane, Poland}
}}
\vspace{2.0cm}
\author{Harmen J. Warringa \\ $\phantom{boktor}$ \\ 
{\it{Department of Physics and Astronomy, Vrije Universiteit,}} \\ 
{\it{De Boelelaan 1081, 1081 HV Amsterdam, The Netherlands}}}
\vspace{1.0cm}

\maketitle

\begin{abstract}
\noindent
The thermodynamics of the $O(N)$ nonlinear sigma model in $1+1$ dimensions 
is studied. We calculate the finite temperature effective potential in leading
order in the $1/N$ expansion and show that at this order the effective potential
can be made finite by temperature independent renormalization.  We will show
that this is not longer possible at next-to-leading order in $1/N$. 
In that case one can only renormalize the minimum of the effective potential
in a temperature independent way, which gives us finite 
physical quantities like the pressure.
\end{abstract}

\thispagestyle{empty}
\newpage

\section{Introduction}
The nonlinear sigma model is a scalar field theory 
with an $O(N)$ 
symmetry. It is described by a Lagrangian density which only consists
 of a kinetic term,
\begin{equation}
      \mathcal{L} = \frac{1}{2} \partial_\mu \phi_i \partial^\mu \phi_i \;,
\end{equation}
and a constraint which enforces all the $\phi$ fields to 
lie on a $N-1$ sphere:
\begin{equation}
   \phi_i(x) \phi_i(x) = N/g^2 \;\;\;\;\; i = 1 \ldots N \;.
\end{equation}
This model has some nice features in $1+1$ dimensions, which makes 
it interesting to study as a toy model for QCD. 
First it is renormalizable. Furthermore it is asymptotically free, 
such that at very high temperatures it approaches a
free field theory. The model also has a dynamically generated mass for 
the $\phi$ fields. If $N=3$ the model has instanton solutions.
Finally, for $N=2$ we recover a free field theory, which
can be used as a check of the calculations.

In this article we will study the
thermodynamical properties of the nonlinear sigma
model. In particular we will calculate the pressure.
In~\secref{sec:pressure} we briefly discuss some aspects of thermal field
theory.
In~\secref{sec:pertexp} we calculate 
the pressure in the weak-coupling expansion.
In~\secref{sec:oneovernexp}, we calculate the
effective potential and pressure to
leading order in the $1/N$ expansion.
The next-to-leading order (NLO) correction
is discussed in \secref{sec:nlopressure}.

\section{The pressure in a field theory}
\label{sec:pressure}
In this section we briefly review how one calculates
the pressure in a thermal field theory. For a more complete
introduction see \citerefs{Kapusta, LeBellac}.

In classical statistical mechanics one can derive all thermodynamic
quantities from 
the partition function. The partition function $\mathcal{Z}$ is given by
\begin{equation}
\mathcal{Z} = \sum_n \left< n \left \vert {\exp[-\beta \hat H]} \right \vert n 
\right > \;,
\end{equation}
where the sum is over all eigenstates of the Hamiltonian $\hat H$
and $\beta = 1 / T$, the inverse temperature. 
For example the pressure ${\cal P}$ is given by
\begin{equation}
{\cal P} = \frac{1}{\beta} \frac{\partial \log \mathcal{Z}}{\partial V} \;.
\end{equation}
We next express the partition function in terms of 
fields. The easiest way to do this is to consider a transition
matrix element in ordinary field theory. One can write
such a transition element in terms of a path integral 
in the following way
\begin{equation}
  \left < \phi_f \left \vert \exp[-i(t_f-t_i) \hat H] \right \vert
  \phi_i \right>
 =  \int \mathcal{D} \phi \; 
       \exp \left [i {\int_{t_i}^{t_f} \measure{t}}
       { \int \measure{^d x}} 
         \mathcal{L(\phi)} \right]  \;,
\end{equation} 
where $\mathcal{L}$ is a Lagrangian density which has a 
Minkowskian metric and does not have derivative interactions. 
Now if one makes the identification $t = -i \tau$ one finds
\begin{equation}
  \left < \phi_f \left \vert \exp[-\beta \hat H] \right \vert
  \phi_i \right>
 =  \int \mathcal{D} \phi \; 
       \exp \left [- {\int_{0}^{\beta} \measure{\tau}}
       { \int \measure{^d x}} 
         \mathcal{L(\phi)} \right]  \;,
\end{equation}
where we from now on denote the zero component of a ($d+1$)-vector
by $\tau$ and hence use a Euclidean metric.
The last equation enables us to write the partition function
in terms of a path integral,
\begin{equation}
  \mathcal{Z}
 =  \int 
    \mathcal{D} \phi \; 
       \exp \left [- \int_{0}^{\beta} \measure{\tau}
      \int \measure{^d x} 
         \mathcal{L(\phi)} \right]_{\phi(\tau = 0) 
  = \phi(\tau = \beta)} \;,
\end{equation}
where one implicitly integrates over all states which obey
the periodicity condition $\phi(\tau = 0, \vec x) = \phi(\tau = \beta, \vec x)$.
So equilibrium thermal field theory is in essence a Euclidean field theory
where one dimension ($\tau$) is compactified to a circle. 
As
a consequence, the Fourier transform of a field 
becomes a sum over
modes, 
\begin{equation}
\phi(\tau, \vec x) = {\frac{1}{\beta} \sum_{n}}
     {\int \spaceint{d}{k}} 
    e^{i \omega_n \tau + i \vec k \cdot \vec x} \tilde \phi(k)
     \equiv \sumint_K  
e^{i \omega_n \tau + i \vec k \cdot \vec x} \tilde \phi(k)\;,
\end{equation} 
where $\omega_n = 2\pi n T$. This implies that in a 
loop diagram one should not take the integral over internal 
momentum but rather the sum-integral $\Sigma \!\!\!\!\! \int$.

Now for example the partition function of the nonlinear sigma model
is given by
\begin{equation}
  \mathcal{Z} =  \int \prod_{i=1}^N \mathcal{D} \phi_i 
         \; \prod_x \delta(\phi_i(x) \phi_i(x) - N / g^2)
         \exp \left [- \int_0^{\beta} \measure{\tau} \int \measure{x}
             \mathcal{L(\phi)} \right] \;,
\end{equation}
where from now on we work in one spatial dimension. 
To obtain the pressure we have to calculate $\mathcal{Z}$. We will 
follow two paths. The first one is making an expansion around 
$g^2 = 0$. This will only give us the leading term of
the pressure.
The second way is an expansion in $1/N$ which will generates additional
contributions which are non-analytical in $g^2$.

\section{The pressure in the weak-coupling expansion}
\label{sec:pertexp}
One can get rid of the constraint by integrating out one 
of the $\phi$ fields, which results in
\begin{equation}
  \mathcal{Z} =  \int \prod_{i=1}^{N-1} \mathcal{D} \pi_i 
  \prod_x \theta\left(N/g^2 - \pi_i \pi_i\right)
  \exp \left [- \int_0^{\beta} \measure{\tau} \int \measure{x} 
  \mathcal{L}_{\mathrm{eff}}(\pi) \right] \;,
\end{equation}
where $\theta(x)$ is the step function and the effective Lagrangian 
density $\mathcal{L}_{\mathrm{eff}}$ is given by
\begin{equation}
 \mathcal{L}_{\mathrm{eff}}(\pi) = \frac{1}{2} \partial_\mu \pi_i 
        \partial^\mu \pi_i
  + \frac{g^2}{2} \frac{(\pi_i \partial_\mu \pi_i)^2}{N  - g^2  \pi_i 
        \pi_i} - \frac{1}{2} \beta V \log
   \left(N/g^2 -  \pi_i \pi_i \right)\;.
\end{equation}
For small values of $g^2$ the $\theta(x)$ function is only
vanishing when $\pi(x)$ is large. Since large values of $\pi$
give a small contribution to the path integral we approximate 
$ \theta(N / g^2 - \pi_i \pi_i) \approx 1$ which gives 
\begin{equation}
  \mathcal{Z} = \int \prod_{i=1}^{N-1} \mathcal{D} \pi_i 
  \exp \left [- \int_0^{\beta} \measure{\tau} \int \measure{^d x} 
  \mathcal{L}_{\mathrm{eff}}(\pi) \right] \;.
\end{equation}

We will not calculate $\cal{Z}$ but rather 
$\frac{1}{\beta V} \log \cal{Z}$, where $V$ is the
volume of our $1$ dimensional space. 
Because $\log \cal{Z}$ is an extensive quantity, 
\ie it is linear in $V$, the
pressure is equal to $\frac{1}{\beta V} \log \cal{Z}$.
Since in general $\frac{1}{\beta V} \log \cal{Z}$ 
does not vanish at zero temperature, we subtract
the zero temperature contribution to normalize the 
pressure to zero at zero temperature.

If $g^2=0$ it can be seen from $\mathcal{L}_{\mathrm{eff}}$ that
one has $N-1$ noninteracting $\pi$ fields. Hence it is easy to
show that leading term is equal to the pressure of $N-1$ free fields
\begin{equation}
   \mathcal{P} = -\frac{N-1}{2} \left[\sumint_K \log(K^2)
  - \int_K \log(K^2) \right]
= (N-1)\frac{\pi}{6}T^2 \;,
\label{eq:freepressure}
\end{equation}
where $K = (\omega_n, k\,)$ is a Euclidean two vector and
we defined
\begin{equation}
  \int_K \equiv \int \spaceint{2}{k} \;.
\end{equation}

By calculating the loop diagrams, 
one can show that up to and including order $g^4$ 
one only finds the pressure of a free gas
in $d = 1 + 1$ 
\cite{Dine, future}. 
However one finds corrections to the free pressure
in a
 $1/N$ expansion. This may indicate that
the pressure is completely 
non-analytical in $g^2$.

\section{The effective potential in leading order in $1/N$}
\label{sec:oneovernexp}
Another way to implement the constraint on the $\phi$ fields is by
using a Lagrange multiplier field which we will denote by $\alpha$. 
This gives the following
expression for the partition function,
\begin{multline}
  \mathcal{Z} =  \int \prod_{i=1}^N \mathcal{D} \phi_i \mathcal{D} \alpha
  \exp \left\{- \frac{1}{2} \int_0^{\beta} \measure{\tau} \int \measure{x} 
   \partial_\mu \phi_i \partial^\mu \phi_i \right.  \\ \left.
   - \frac{1}{2} \int_0^{\beta} \measure{\tau} \int \measure{x} 
  \alpha(x) [\phi_i(x) \phi_i(x) - N /g^2] \right \} \;.        
\end{multline}
In this way
the action still is quadratic in the $\phi$ fields, so one can
easily integrate them out. This gives
\begin{equation}
   \mathcal{Z} =  \int \mathcal{D} \alpha
    \exp \left \{
- S[\alpha]
     + \frac{N}{2 g^2} \int_0^{\beta} \measure{\tau} \int \measure{x}
       \alpha(x) \right \} \;,
\end{equation}
where
\begin{equation} 
  S[\alpha] = \frac{N}{2} \mathrm{Tr} \log[ -\partial^2 + \alpha(x)] \;.
\end{equation}

The pressure is equal to the minimum of the effective potential, which one
can calculate by expanding the $\alpha$ field around its vacuum
expectation value $m^2$. By considering the propagator of the $\phi$ 
fields, one can show that to leading order in $1/N$, $m$ is equal
to the physical mass of the $\phi$ fields. This is, however, not
longer the case at NLO, \cite{Flyvbjerg, Andersen:2003va}.
The effective potential can be obtained
from the effective action by division by $\beta V$. 
To calculate the effective potential we
write 
$\alpha = m^2 + \tilde \alpha / \sqrt{N}$ 
and expand the action around $m^2$ \cite{Novikov},
\begin{multline} 
  S[\alpha] = \frac{N}{2} \mathrm{Tr} \log[ -\partial^2 + m^2 ] 
        + \frac{\sqrt{N}}{2} \mathrm{Tr} \left (
   \frac{1} { -\partial^2 + m^2} \tilde \alpha \right)
  + \frac{1}{4} \mathrm{Tr} 
  \left(\frac{1} {-\partial^2 + m^2 } 
  \tilde \alpha \right)^2
  + \mathcal{O}( 1 / \sqrt{N} ) \;.
\label{eq:effactionexpansion}
\end{multline}
From this equation it can easily be seen that
the effective potential can be calculated in a $1/N$ expansion.
The leading order effective potential is given by the classical
action. The corrections are obtained by integrating over the
$\tilde \alpha$ field.

To calculate the leading order effective potential
we introduce a momentum cutoff $\Lambda$ and
subtract $m$ and $T$-independent constants from
the effective potential. This subtraction will not
change the physics, since it only shifts the whole
effective potential by a constant.
One finds for the effective potential at leading order in $1/N$ 
\begin{eqnarray}
  \mathcal{V}(m^2) &=& 
  \frac{N m^2}{2 g_b^2}
  -\frac{N}{2} \left [ \sumint_P \log(P^2 + m^2) - \int_P \log(P^2)
        \right ] 
  \\
  &=& 
  \frac{N}{2} \left [
  \frac{m^2}{g_b^2}
  - \frac{m^2}{4\pi} \left(1 +  \log \frac{\Lambda^2}{m^2} \right)
  + \frac{T^2}{4\pi} J_0(\beta m)  \right]
  \;,
\end{eqnarray}  
where $g_b$ is the bare coupling constant. $J_0(\beta m)$
is given by
\begin{equation}
   J_0(\beta m) 
     = \frac{8}{T^2} \int_{0}^{\infty} \measure{p} 
  \frac{p^2 n(\omega_p)}{\omega_p} \;,
 \label{eq:jzero}
\end{equation}
where $n(x) = 1 / (e^x - 1)$ and $\omega_p^2 = p^2 + m^2$.
One is able to renormalize the 
leading order effective potential in a temperature
independent way by replacing $g^2_b \rightarrow Z_{g^2} g^2(\mu)$ where
\begin{equation}
 \frac{1}{Z_{g^2}} = 1 + \frac{g^2}{4\pi} \log \frac{\Lambda^2}{\mu^2} \;.
\end{equation}
and $g^2 = g^2(\mu)$.
From this equation it follows that the $\beta$-function of $g^2$ is given
by
\begin{equation}
  \beta(g^2) \equiv \mu
   \frac{\mathrm{d} g^2(\mu)}{\mathrm{d}\mu}
  = - \frac{g^4 (\mu)}{2 \pi} \;.
\end{equation}
The leading order $\beta$-function is exact in $g^2$.
Since the $\beta$-function is negative, $g^2$ approaches 
zero for large values of $\mu$. This shows that the 
theory is asymptotically free.

With use of the renormalization of the coupling constant
one finds the following finite expression for the effective potential
\begin{equation}
  \mathcal{V}(m^2) =
  \frac{N}{2} \left[ \frac{m^2}{g^2}
  -\frac{m^2}{4\pi}
  \left( 1 +  
  \log \frac{\mu^2}{m^2} \right)
  + \frac{1}{4\pi} T^2 J_0(\beta m)  
  \right ]\;.
\end{equation}
One can easily show that the effective potential is 
independent of the renormalization scale $\mu$. This is expected since
the choice of $\mu$ is completely arbitrary.

\begin{figure}[htb]
\begin{center}
\scalebox{0.8}{\includegraphics{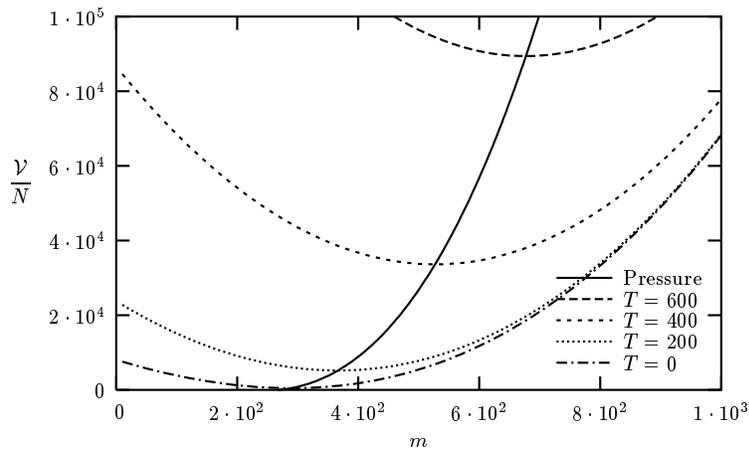}}
\caption{The leading order effective potential 
     as function of $m$ for different temperatures with
$g^2(\mu = 500) = 10$.}
\label{fig:effpot}
\end{center}
\end{figure}

To obtain the pressure, one has to minimize
the effective potential with respect to $m^2$.
Minimization gives the so-called gap equation
\begin{equation}
 \frac{1}{g^2} = \sumint_P \frac{1}{P^2 + m^2} = 
 \frac{1}{4\pi} \log \left(\frac{\mu^2}{m^2} \right) 
 + \frac{1}{4\pi} J_1(\beta m) 
  \equiv \frac{1}{4 \pi} \log \left(\frac{\mu^2}{\bar m^2} \right)  \;, 
  \label{eq:gapequation}
\end{equation}
where $J_1(\beta m)$ is defined by
\begin{equation}
  J_1(\beta m)
  = 4 \int_{0}^{\infty} \measure{p} \frac{n(\omega_p)}{\omega_p} 
 \label{eq:jone} \;.
\end{equation} 
The solution of the gap equation determines the leading
order physical mass of the $\phi$ fields as a function of 
temperature. At $T=0$ one can solve this equation to show that
the mass is completely non-analytical in $g^2$,
\begin{equation}
  m_{T=0} = \mu \exp \left(-\frac{2 \pi}{g^2} \right) \;.
\label{mt0}
\end{equation}
We can use \eqref{mt0} to normalize the
minimum of the effective potential at $T=0$ to be zero which gives
\begin{equation}
  \mathcal{V}(m^2) =
  \frac{N}{2} \left[ \frac{m^2}{g^2}
  -\frac{m^2}{4\pi}
  \left( 1 +  
  \log \frac{\mu^2}{m^2} \right)
  + \frac{1}{4\pi} T^2 J_0(\beta m)  
  + \frac{m^2_{T=0}}{4\pi} \right ]\;.
\end{equation}
The effective potential 
as a function of $m$ for different temperatures
is shown in~\figref{fig:effpot}, 
for the arbitrary choice $g^2(\mu = 500) = 10$.
The quantities $T$, $m$, $\mu$, $\mathcal{V} / T$ and
$\mathcal{P} / T$ are all in 
the same arbitrary units.
The solid curve which is the minimum of the effective potential
is equal to the pressure.

\section{Next-to-leading order correction in $1/N$}
\label{sec:nlopressure}
\begin{figure}[htb]
\begin{center}
\scalebox{0.8}{\includegraphics{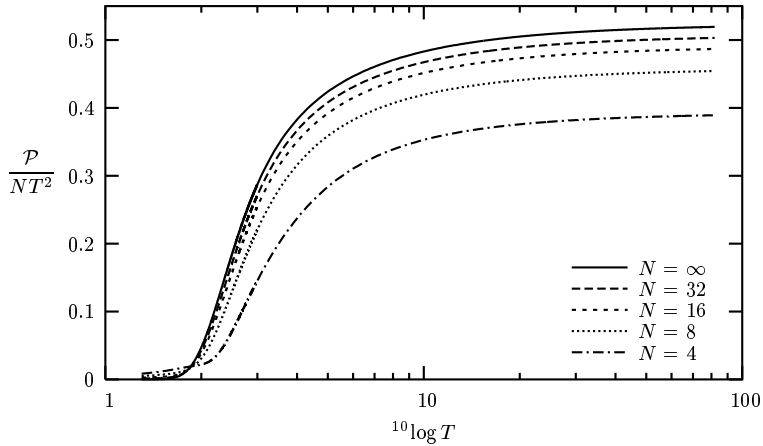}}
\caption{Pressure $\cal{P}$ normalized to $NT^2$ as a function of 
         temperature for different values of $N$ with $g^2(\mu = 500)
         = 10$  
         \cite{Andersen:2003va}. }
\label{fig:pressure}
\end{center}
\end{figure}

The term linear in $\alpha$ in \eqref{eq:effactionexpansion}
gives no contribution to the effective potential since
it gives rise to a tadpole \cite{Zinn-Justin}.
The first $1/N$ correction to the effective potential stems
from the last term of \eqref{eq:effactionexpansion}. By
going to momentum space one can show that the correction is
given by
\begin{equation}
      \mathcal{V}_1(m^2) 
  = - \frac{1}{2} 
  \sumint_P \log\left[ \sumint_Q \frac{1}{Q^2 + m^2}
         \frac{1}{(P+Q)^2 + m^2} \right] \;.
\end{equation}
We calculated this correction in Ref.~\cite{Andersen:2003va}.
In the limit $\Lambda \rightarrow \infty$, one obtains
\begin{multline}
  \mathcal{V}_1(m^2) 
  =  
  -\frac{1}{8 \pi} \left(\Lambda^2 \ln \ln \frac{\Lambda^2}{\bar m^2}
 - \bar m^2 \, \mathrm{li}\, \frac{\Lambda^2}{\bar m^2} \right) 
 - \frac{m^2}{4 \pi} 
  \left ( \ln \ln \frac{\Lambda^2}{\bar m^2} -
  \ln \frac{\Lambda^2}{4 m^2} \right) +
  F(m, T)\;,
 \label{eq:nloeffpot}
\end{multline} 
where $\bar m$ is defined in \eqref{eq:gapequation}.
In~(\ref{eq:nloeffpot}), we have subtracted $m$ and $T$-independent 
constants and dropped
terms that vanish in the limit $\Lambda \rightarrow \infty$.
$F(m, T)$ is a finite term and the logarithmic integral $\mathrm{li}$ 
is defined by
\begin{equation}
  \mathrm{li}(x) = \mathcal{P} \int_{0}^{x} \measure{t} \frac{1}{\log t}\;,
\end{equation}
where $\cal{P}$ stands for principal value. The first
two terms of \eqref{eq:nloeffpot} are problematic. It is 
impossible to remove these divergences 
by renormalizing $g^2$ in a temperature independent way
or by subtracting $m$ and $T$-independent
constants. However this is possible at the 
minimum of the effective potential. 
At the minimum, one can use the leading order gap 
equation, \eqref{eq:gapequation}, to show
that $\bar m$ is independent of $T$. So one could add
\begin{equation}
  \frac{\Lambda^2}{8 \pi} \left \{ \ln \frac{4\pi}{g_b^2}
 - \exp\left(-\frac{4\pi}{g_b^2} \right)
  \mathrm{li}\, 
  \left[\exp \left(\frac{4\pi}{g_b^2}\right) \right] \right \} 
\end{equation}
to the effective potential which yields an effective potential 
that can be renormalized at the minimum. 
Using this renormalization at the minimum we have calculated 
the pressure $\cal{P}$ as a function of $N$. The result 
is depicted in \figref{fig:pressure}. One clearly sees
a crossover which is not a phase transition.
This is in accordance with the Mermin-Wagner-Coleman theorem
\cite{Mermin-Wagner, coleman}
which
forbids spontaneous breakdown of a continuous symmetry in $1+1$ dimensions.  
The figure furthermore shows that the
$1/N$ expansion is relatively good since
the corrections are really of order $1/N$. 
Finally it can be seen from the figure
that the theory is asymptotically free, 
because in the limit $T \rightarrow \infty$
the pressure approaches the pressure of a free gas, \eqref{eq:freepressure}.

\section{Summary and Conclusions}
\label{sec:conclusions}
We find that the pressure of the nonlinear sigma model in the weak-coupling
expansion through order $g^4$ only consist of the free term.
Furthermore, we showed that in a $1/N$ expansion we can renormalize the 
leading order effective
potential in a temperature independent way. This is, however, impossible
for the effective potential at next-to-leading order in $1/N$. In
that case one can only renormalize in a temperature-independent 
way a physical quantity,
like the pressure. 

\section*{Acknowledgments}
This work has been carried out in collaboration with Jens O. Andersen
and Dani\"el Boer.

\newpage
\bibliographystyle{unsrt}

\begin{thebibliography}{99}

\bibitem{Kapusta}
J.~I.~Kapusta,
{\it Finite-temperature field theory},
Cambridge University Press (1989)

\bibitem{LeBellac}
M.~Le~Bellac,
{\it Thermal field theory},  
Cambridge University Press, (2000)

\bibitem{Dine}
M.~Dine and W.~Fischler,
Phys.\ Lett.\ B {\bf 105}, 207 (1981).

\bibitem{future}
J.~O. Andersen, D. Boer and H.~J. Warringa, in preparation.

\bibitem{Flyvbjerg}
H.~Flyvbjerg,
Phys.\ Lett.\ B {\bf 245}, 533 (1990).

\bibitem{Andersen:2003va}
J.~O.~Andersen, D.~Boer and H.~J.~Warringa,
arXiv:hep-ph/0309091.

\bibitem{Novikov}
V.~A.~Novikov, M.~A.~Shifman, A.~I.~Vainshtein and V.~I.~Zakharov,
Phys.\ Rept.\  {\bf 116}, 103 (1984).

\bibitem{Zinn-Justin}
J.~Zinn-Justin, {\it Quantum Field Theory And Critical Phenomena}, Oxford
University press (1996), 3rd edition.

\bibitem{Mermin-Wagner}
N.~D.~Mermin and H.~Wagner,
Phys.\ Rev.\ Lett.\  {\bf 17}, 1133 (1966).

\bibitem{coleman}
S.~R.~Coleman,
Commun.\ Math.\ Phys.\  {\bf 31}, 259 (1973).

\end{thebibliography}

\end{document}